\documentstyle[multicol,prb,aps,epsfig]{revtex}
\begin{document}
\draft
\title{Non-Gaussian Resistance Fluctuations in Disordered Materials} 
	
\author{C. Pennetta$^{1}$, E. Alfinito$^{1}$, L. Reggiani$^{1}$, 
S. Ruffo$^{2}$}

\address{$^1$ INFM - National Nanotechnology Laboratory
and Dipartimento di Ingegneria dell'Innovazione,
\\Universit\`a di Lecce, Italy, Via Arnesano, I-73100, Lecce, Italy, \\
$2$ CSDC and Dipartimento di Energetica ``Sergio Stecco'', INFN and INFM,
\\Universit\`a di Firenze, Via S. Marta, 3, Firenze, 50139, Italy.
\thanks{Corresponding authors e-mail: cecilia.pennetta@unile.it }}
%
%
\maketitle
\begin{abstract}
We study the distribution of resistance fluctuations of conducting thin films 
with different levels of internal disorder. The film is modeled as a resistor 
network in a steady state determined by the competition between two biased 
processes, breaking and recovery of the elementary resistors. The fluctuations
of the film resistance are calculated by Monte Carlo simulations which are 
performed under different bias conditions, from the linear regime up to the 
threshold for electrical breakdown. Depending on the value of the external
current, on the level of disorder and on the size of the system, the 
distribution of the resistance fluctuations can exhibit significant deviations
from Gaussianity. As a general trend, a size dependent, non universal 
distribution is found for systems with low and intermediate disorder. 
However, for strongly disordered systems, close to the critical point of the 
conductor-insulator transition, the non-Gaussianity persists 
when the size is increased and the distribution of resistance fluctuations is 
well described by the universal Bramwell-Holdsworth-Pinton distribution. 
\end{abstract}
\vspace{0.7cm}
\centerline{Keywords:Non-Gaussian distributions, Disordered materials, 
Non-equilibrium steady states, Electrical breakdown}
\vspace{0.7cm}
%
%
\begin{multicols} {2}

\section{INTRODUCTION}
In a macroscopic system, the fluctuations of a global quantity are generally 
expected to follow a Gaussian distribution around the average value. This
expectation relies on the assumption, frequently true, that the 
system can be divided into statistically independent microscopic or mesoscopic
elements. For such a system, the conditions of validity of the central-limit 
theorem are in fact satisfied and, consequently, Gaussian distributions of 
fluctuations are actually observed. On the other hand, when the correlations 
between the different elements of the system are significant, the 
central-limit theorem is no longer valid and a non-Gaussian distribution of 
fluctuations is expected. Therefore, non-Gaussian fluctuation distributions 
of global quantities are the signature of a correlated system. As correlations
become important near the critical points of phase transitions, non-Gaussian 
distributions are usually observed near criticality
\cite{chakrabarti,weissman,bramwell_nat,bramwell_prl,aji,antal,eisler}. 
In these conditions, the self-similarity of the system over all the
scales, from a characteristic microscopic length up to the size of the system 
(which for a system of finite size sets the cut-off for the correlation 
length), has important implications on the fluctuation distribution
\cite{chakrabarti,weissman,bramwell_nat,bramwell_prl,bramwell_pre,aji,antal,eisler,racz}. 
Recently, new light has been shed on this subject by the remarkable 
discovery made by Bramwell, Holdsworth and Pinton (BHP) \cite{bramwell_nat}
of a common behavior of the distribution of fluctuations of two quite different
systems. Namely, the distribution of the fluctuations of the power consumption 
measured in confined turbulent-flow experiments and the distribution of 
magnetization fluctuations in a two-dimensional XY model in the spin-wave 
regime at low temperature  \cite{bramwell_nat}. Since these two systems appear
to have very little in common, BHP advanced the reasonable suggestion that the
origin of this common distribution should be attributed to scale invariance, 
the only property apparently shared by the two systems \cite{bramwell_nat}. 
This suggestion has been supported by the subsequent finding that many scale 
invariant systems, in both equilibrium and non-equilibrium steady states, 
display the same functional form for the distribution of fluctuations 
\cite{bramwell_prl,bramwell_pre,portelli,jensen,aji}. 
Concerning this point, it must be noted that the probability density function 
(PDF) of the distribution of fluctuations of a critical system can be 
considered as universal, in the sense that, when properly normalized, it 
depends on very few basic symmetries that define the universality class of the
system \cite{bramwell_pre}. Thus, the BHP distribution describes a universal
behavior of the fluctuations of a critical system. In any case, it must be 
remarked that not all critical systems fluctuate according to this 
distribution \cite{aji,antal,eisler,jensen}. On the other hand, far from 
criticality, the correlations between different elements of the systems can 
also be important. This is particularly true for systems in non-equilibrium 
steady-states, where non-Gaussian fluctuations are frequently observed 
\cite{chakrabarti,weissman,eisler,ausloos}. Therefore the study of 
non-Gaussian fluctuations and the understanding of their link with other 
features of the system can provide new insights on some basic properties 
of complex systems
\cite{chakrabarti,weissman,bramwell_prl,aji,antal,eisler,jensen,racz}. 

Here, we study the distribution of resistance fluctuations of conducting
thin films of different size and with different levels of internal disorder. 
The resistance fluctuations are investigated under different bias conditions, 
from the linear response regime up to the threshold for electrical breakdown. 
This last phenomenon, which consists of an irreversible increase of the 
resistance, is thus associated with a conductor-insulator transition and it 
occurs in conducting materials stressed by high current densities 
\cite{hansen,sornette,andersen,stan_zap,bardhan,bardhan1,prl_fail}. In
our study we make use of the Stationary and Biased Resistor Network (SBRN) 
model \cite{upon99,SBRN_1,SBRN_2}. This model provides a good description of 
many features associated with the electrical instability of composites 
materials \cite{bardhan,bardhan1} and with the electromigration damage of 
metal lines \cite{prl_fail,pen_em}, two important classes of breakdown 
phenomena. The film is modeled as a resistor network which reaches a steady 
state determined by the competition between two biased stochastic processes, 
breaking and recovery of the elementary resistors. The resistance and its 
fluctuations are then calculated by Monte Carlo simulations. Resistance 
fluctuations are found to deviate from Gaussianity near electrical breakdown. 
As a general trend, a non universal distribution of fluctuations is found for 
systems with low and intermediate disorder, where the deviations from 
Gaussianity vanish in the large size limit. However, for highly 
disordered systems, close to the critical point of the conductor-insulator 
transition, these deviations from Gaussianity persist when the size is 
increased and the distribution of resistance fluctuations is well described by
the universal BHP distribution \cite{bramwell_nat,bramwell_prl}. 

\section{MODEL}
We describe a conducting film with granular structure as a two-dimensional 
resistor network. Precisely, we consider a square-lattice of $N \times N$ 
resistors, where $N$ determines the linear size of the network. This lattice 
lies on an insulating substrate at a given temperature $T_0$, which acts 
as a thermal bath. Each resistor can be in two different states
\cite{hansen,gingl,prl_fail}: 
(i) regular, corresponding to a resistance $r_n =r_0[1+\alpha(T_n - T_0)]$, 
and (ii) broken, corresponding to an effectively ``infinite'' resistance, 
$r_{OP} = 10^9 r_n$ (resistors in this state will be called defects). In the 
previous expression of $r_n$, $\alpha$ is the temperature coefficient of the 
resistance and $T_n$ the local temperature. This latter is determined by Joule
heating effects\cite{hansen,gingl,prl_fail} and thermal exchanges 
between neighbor resistors \cite{prl_fail}:
\begin{equation}
T_{n}=T_{0} + A [ r_{n} i_{n}^{2} + (3/4N_{neig}) 
\sum_{m=1}^{N_{neig}} ( r_{l} i_{l}^2   - r_n i_n^2)]
\end{equation} 
where, $i_{n}$ is the current flowing in the n{\em th} resistor and $N_{neig}$
the number of nearest neighbors over which the summation is performed. The
parameter $A$ represents the thermal resistance of each resistor and  
sets the importance of Joule heating effects. By taking the above 
expression for $T_n$ we are assuming an instantaneous thermalization 
\cite{prl_fail} of each resistor at the value $T_n$. The external bias 
consists of a constant current $I$ applied through perfectly conducting bars 
at the left and right sides of the network. In the initial state of the 
network (no external bias) all the resistors are identical: $r_n \equiv r_0$. 

We assume that two competing biased processes act to determine the evolution 
of the network \cite{upon99,SBRN_1,SBRN_2}. These two processes consist of 
stochastic transitions between the two possible states of each resistor and 
they are taken to occur through thermal activation, with probabilities 
\cite{gingl}: $W_{Dn}=exp[-E_D/k_B T_n]$ and $W_{Rn}=exp[- E_R/k_B T_n]$, 
characterized by the two activation energies, $E_D$ and $E_R$ ($k_B$ being the
Boltzmann constant). The time evolution of the network is obtained by 
Monte Carlo simulations which update the network resistance after breaking and 
recovery processes, according to an iterative procedure described in details 
in Ref. \cite{SBRN_1} The sequence of successive configurations provides a 
resistance signal, $R(t)$, after an appropriate calibration of the time scale. 
Then, depending on the stress conditions ($I$ and $T_0$) and on the network 
parameters (size, activation energies and other parameters related to the 
material like $r_0$ and $\alpha$), the network either reaches a steady state 
or undergoes an irreversible electrical failure \cite{SBRN_1,SBRN_2}. This 
latter possibility is associated with the achievement of the percolation 
threshold, $p_c$, for the fraction of broken resistors \cite{stauffer}. 
Therefore, for a given network at a given temperature, a threshold current 
value, $I_B$, exists above which electrical breakdown occurs \cite{SBRN_1}. 
For values of the current below this threshold, the steady state of the 
network is characterized by fluctuations of the fraction of broken resistors, 
$\delta p$, and of the resistance, $\delta R$, around their respective average
values $<p>$ and $<R>$. In particular, we underline that in the vanishing 
current limit (random percolation) \cite{prl_stat}, the ratio 
$(E_D -E_R)/k_B T_0$ determines the average fraction of defects and thus the 
level of disorder inside the network. In the following we analyze the results 
of simulations performed by considering networks of different sizes, with 
different levels of disorder and stressed by different currents at room 
temperature, $T_0=300$ (K). In all the cases we take $E_D = 0.170$ (eV), 
$r_0=1$ ($\Omega$), $\alpha = 10^{-3}$ (K$^{-1}$), $A=5 \times 10^5$ (K/W) 
(these values are chosen as physically reasonable). The values of $N$ range 
between $100 \div 150$, while $E_R$ between $0.026 \div 0.164$ (eV). 

\section{RESULTS}
The resistance of a $100 \times 100$ network is reported as a function of the
time in Fig. 1. The different curves, starting from the bottom, are obtained 
for increasing values of the external current. In particular, all the 
evolutions in Fig. 1 are obtained by taking the activation energy of the 
recovery process equal to $E_R = 0.103$ eV, a value which leads to a network
with an intermediate level of disorder. The first three curves are associated
with steady states of the network. More precisely, the lowest curve 
corresponds to the linear response regime and it is obtained for 
$I=0.0013 A < I_0$, where $I_0$ is the current value associated with the onset
of the nonlinearity of the I-V characteristic \cite{SBRN_1}. The second curve 
corresponds to the nonlinear regime and it is obtained for $I=0.70$ A. The 
third corresponds to the threshold for electrical breakdown which occurs for 
$I>I_B= 0.95 (A)$. By contrast, the highest curve displays the resistance 
evolution of a network undergoing electrical breakdown and it is obtained for 
a current $I=1.05 (A)$. Overall, Fig. 1 illustrates qualitatively two 
important features of the electrical response of a conducting film. First, it 
evidences that the linear regime occurring for $I<I_0$ is followed by a 
nonlinear regime where the average resistance increases significantly at
increasing current \cite{SBRN_1,bardhan}. Second, it shows that the amplitude 
of the resistance fluctuations increases strongly with the external bias. This
amplification of the fluctuations becomes particularly important when the 
current reaches the threshold for breakdown (third curve from the bottom in 
Fig. 1). A detailed analysis of the behavior of the average resistance and of 
the relative variance of resistance fluctuations as a function of the current 
can be found in Refs. \cite{SBRN_1,SBRN_2}

\begin{figure}
   \epsfig{figure=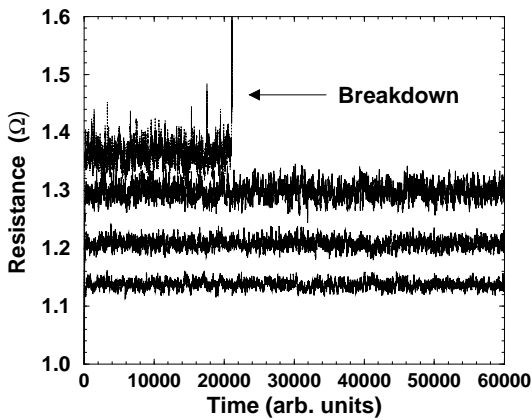,width=7.0cm}
   \caption{Resistance evolutions of a $100 \times 100$ network stressed by 
   increasing current values. Starting from the bottom: $I=0.0013$ A (linear 
   regime), $I=0.70$ A (nonlinear regime), $I=0.95$ A (threshold current), 
   $I=1.05$ A (breakdown). The recovery energy is $E_R=0.103$ eV and 
   corresponds to an intermediate level of disorder. } 
   {\label{fig:1}}
   \end{figure} 

The distributions of the resistance fluctuations, $\delta R$, for the three 
steady state signals in Fig. 1 are reported in Fig. 2. Precisely, by denoting
with $\Phi$ the PDF of the $\delta R$ distribution and with $\sigma$ the root 
mean square deviation from the average resistance value, we plot in Fig. 2 on 
a lin-log scale the product $\sigma \Phi$ as a function of $(<R>-R)/\sigma$. 
This normalized representation, by making the distribution independent of its 
first and second moments, is particularly convenient to explore the functional
form of any distribution \cite{bramwell_nat,bramwell_prl}. Here the PDFs have 
been calculated by considering time series containing about $1.2 \times 10^6$ 
resistance values. In this figure, the data represented by crosses, small 
circles and up triangles are obtained respectively for $I=0.0013, 0.70,
0.95$ A (i.e. linear regime, nonlinear regime and threshold for breakdown). 
For comparison, in this figure we also report with a dashed curve the Gaussian
distribution (which in this normalized representation has zero mean and unit 
variance) and, with a continuous curve, the BHP distribution 
\cite{bramwell_nat,bramwell_prl}. By defining the normalized variable, 
$y\equiv (R-<R>)/\sigma$, the normalized PDF, $\Pi(y)\equiv \sigma \Phi(y)$ 
and $x \equiv b(y-s)$, the BHP distribution has the following functional 
form \cite{bramwell_prl}: 

\begin{equation}
\Pi(y) = K [e^{x - e^{x}}]^a \label{eq:bhp}
\end{equation} 

where $a=\pi/2$, $b=0.936 \pm 0.002$, $s=0.374 \pm 0.001$ and 
$K=2.15 \pm0.01$ \cite{bramwell_prl}. This expression can be considered as
a generalization of the Gumbel distribution, which is often associated with 
the occurrence of rare events. Figure 2 shows a remarkable deviation from the 
Gaussian behavior when the network is stressed by the threshold current $I_B$. 
We have found that in this case the PDF is well fitted by the 
Eq.~(\ref{eq:bhp}), once the parameters $a$, $b$, $s$ and $K$ are taken as 
fitting parameters \cite{pen_hcis} In any case, we emphasize the fact that, at
least for a network with intermediate level of disorder, such as that 
considered in Figs. 1 and 2, the deviations from Gaussianity remain weak even 
for current values in the nonlinear regime and become important only close to 
the electrical breakdown.

\begin{figure}
   \epsfig{figure=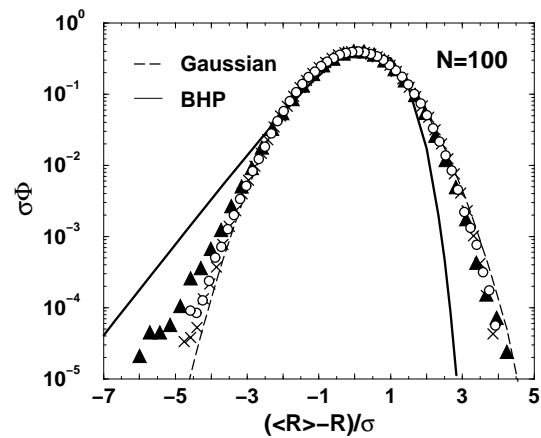,width=7.0cm}
   \caption{Normalized PDF of resistance fluctuations for a network of size 
   $100\times100$ biased by $I=0.0013$ A (crosses, linear regime),
   $I=0.70$ A (small circles, nonlinear regime), $I=0.95$ A (up triangles, 
   threshold current). The recovery energy is the same of Fig. 1. The thick 
   solid curve and the dashed one correspond to the BHP and Gaussian 
   distributions, respectively.}
   {\label{fig:2}}
   \end{figure} 

The effect of the size of the network on the distribution of resistance 
fluctuations, is analyzed in Figs. 3 and 4. Figure 3 compares the PDF 
obtained for a current in the linear regime and applied to a network 
$100 \times 100$ (the data, reported by crosses, are the same of Fig. 2) with  
the PDF obtained for a current in the same regime applied to a network 
$125 \times 125$ (diamonds). The two networks have the same parameters and 
differ only for the size, thus they share the same level of internal disorder. 
Figure 3 shows that in the linear regime the distribution of $\delta R$ 
is Gaussian for all system sizes. In case, a weak non-Gaussian tail  
appears for very small systems that vanishes for systems of larger size. 

\begin{figure}
   \epsfig{figure=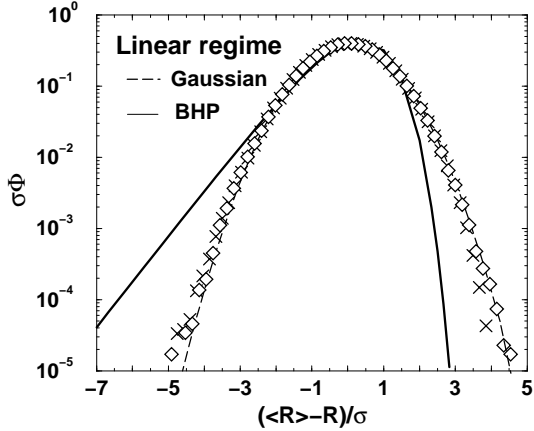,width=7.0cm}
   \caption{Normalized PDF of resistance fluctuations for networks of size 
   $100\times100$ (crosses) and $125\times125$ (diamonds). The networks
   are biased by currents in the linear regime. The recovery energy is 
   the same of Fig. 1. The thick solid curve and the dashed one correspond 
   to the BHP and Gaussian distributions, respectively.} 
   { \label{fig:3}}
   \end{figure} 
\begin{figure}
   \epsfig{figure=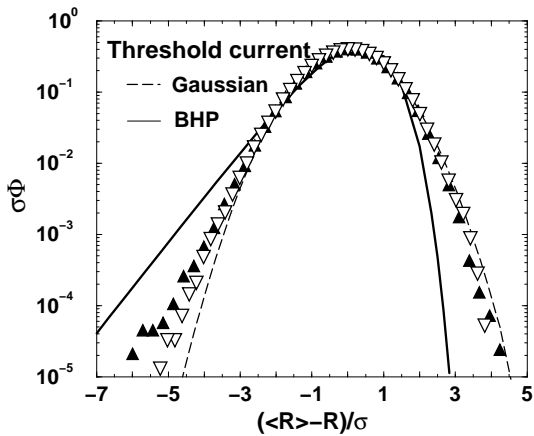,width=7.0cm}
   \caption{Normalized PDF of resistance fluctuations for networks of size 
   $100\times100$ (up triangles) and $125\times125$ (down triangles). The 
   networks are biased by $I=0.95$ A and $I=1.15$ A which correspond to their 
   respective threshold for breakdown. The recovery energy $E_R$ is the same 
   of the previous figures and the thick solid and the dashed curves have the 
   same meaning of Fig. 3.} 
   {\label{fig:4}}
   \end{figure} 

Figure 4 displays a comparison similar to that in Fig. 3 for two networks of 
size $100 \times 100$ (up triangles, same data of Fig. 2) and $125 \times 125$ 
(down triangles), stressed by currents corresponding to the threshold for 
breakdown. The current values are $I=0.95, 1.15$ A, respectively for the
networks with $N=100$ and $N=125$. We can see that in both cases the PDFs 
exhibit non-Gaussian tails. However, the same figure shows that the 
non-Gaussianity is weaker for the system of larger size. By performing a
detailed investigation on systems of different sizes we have found that this 
trend is systematic \cite{physicaa}. Therefore, we conclude that, at least  
when networks with intermediate level of disorder are considered, as in the 
case of Figs. 1-4, the deviations from Gaussianity are related to the finite 
size of the system.

Now, before discussing the role of disorder in the breakdown process,
it is convenient to distinguish between the intrinsic disorder and the 
disorder driven by the external bias. For given values of $T_0$ and $E_D$, 
the average fraction of broken resistors in the vanishing current limit, 
$<p>_0$, is only determined by the recovery energy $E_R$ \cite{prl_stat}. 
This average fraction of defects represents an intrinsic property 
of the system, which characterizes the conducting material and sets 
the level of disorder inside the network in absence of an external bias. 
We note nate the steady state of a network in this vanishing current limit is 
determined by the competition between two random percolations \cite{prl_stat}. 
On the other hand, for $I>I_0$, there is also an additional component 
of disorder due to the defects generated by the external current
\cite{SBRN_1}. Therefore, the average fraction of defects $<p>$ becomes 
dependent on the current and its relative variation, $[<p>-<p>_0]/<p>_0$, 
scales as \cite{SBRN_1} $(I/I_0)^2$. Of course, this dependence of $<p>$
on $I$ is important because it actually drives the system towards the 
breakdown. However, it is also crucial to point out the role of the 
intrinsic disorder on the breakdown process and its effect on the distribution 
of resistance fluctuations. Therefore, in the following we will consider 
networks with different values of the recovery activation energy $E_R$.

\begin{figure}
   \epsfig{figure=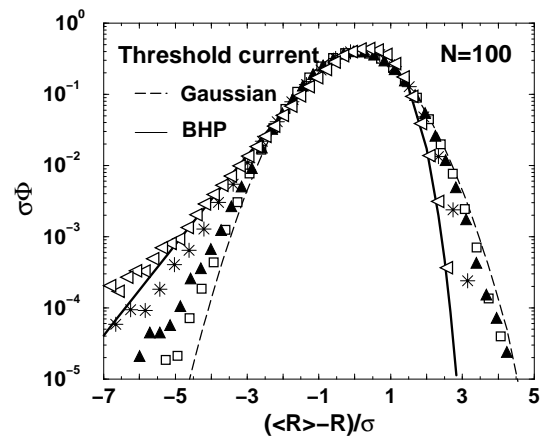,width=7.0cm}
   \caption{Normalized PDF of resistance fluctuations of $100\times 100$ 
    networks at increasing  $E_R$ values (level of disorder). In all the cases
    the current value corresponds to the respective threshold for breakdown. 
    Precisely, squares: $E_R=0.026$, $I_B=3.5$; up triangles: $E_R=0.103$, 
    $I_B=0.95$; stars: $E_R=0.155$, $I_B=0.11$, left triangles:  
    $E_R=0.164$, $I_B=0.009$ (energies in eV, currents in A).} 
   {\label{fig:5}}
   \end{figure} 

In a previous work \cite{upon02} we have shown that for an arbitrary value
of $E_R$ satisfing the stability condition  \cite{prl_stat,SBRN_2}, i.e. 
providing a steady state of the network, the electrical breakdown is associated
with a first order phase transition. This result agrees with the behavior 
observed in electrical breakdown experiments, performed in the Joule regime 
of composites \cite{bardhan,bardhan1}. Nevertheless, it can be shown
\cite{unpub} that when $E_R$ reaches its maximum value compatible with a 
steady state, $E_{R,MAX}$, the conductor-insulator transition becomes of the 
second order. This change in the order of the transition, when going from 
low to high disordered systems, has been also predicted by Andersen 
et al. \cite{sornette} 

Therefore, we report in Fig. 5 the PDFs of resistance fluctuations 
of $100 \times 100$ networks characterized by different values of $E_R$.
The  PDFs have been calculated for current values corresponding to the 
breakdown threshold. We can see that the non-Gaussianity of the distribution 
at $I=I_B$ increases systematically at increasing values of $E_R$ (i.e. at 
increasing level of the intrinsic disorder). In particular, Fig. 5 shows that 
when the value of $E_R$ is very close to $E_{R,MAX}$, and thus the system 
approaches the critical point, the PDF achieves (left triangles) the BHP form.
In these conditions, the PDF should become independent of the system size
\cite{bramwell_nat,bramwell_prl}. 

\begin{figure}
   \epsfig{figure=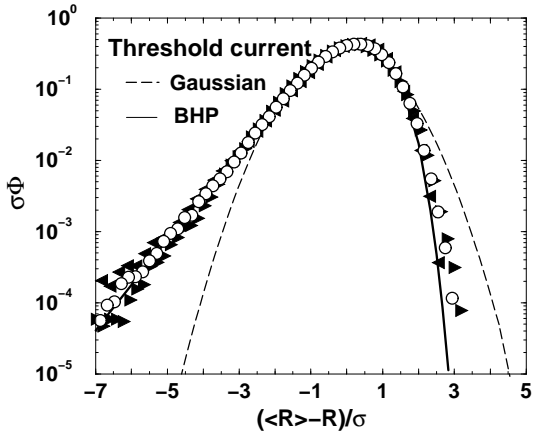,width=7.0cm}
   \caption{Normalized PDF of resistance fluctuations for networks of    
   increasing size and close to the critical conditions: the recovery energy 
   is $E_R=0.164$ eV and the bias current corresponds to the threshold for 
   breakdown. Precisely,  left triangles refer to a network 
   $100\times100$ biased by $I=0.0090$ A, circles to a network $125\times125$ 
   biased by $I=0.0011$ A  and right triangles to a network $150\times150$ and
   $I=0.0013$ A.}
   {\label{fig:6}}
   \end{figure} 

Indeed, Fig. 6 confirms this behavior. 
In this figure we report the PDFs calculated for the three networks of size 
$100 \times 100$ (left triangles, same data of Fig. 5), $125 \times 125$ 
(circles) and $150 \times 150$ (right triangles). In all the cases the 
recovery energy is $E_R=0.164 eV \approx E_{R,MAX}$ and the external current
corresponds to the threshold value for breakdown ($I_B=0.0090, 0.011, 0.013$ A,
respectively). The data in Fig. 6 shows that, when $E_R$ is very close to 
$E_{R,MAX}$ and thus the system approaches the critical point, the PDFs of 
resistance fluctuations become independent of the system size and they are 
well described by the BHP distribution. 

\begin{figure}
   \epsfig{figure=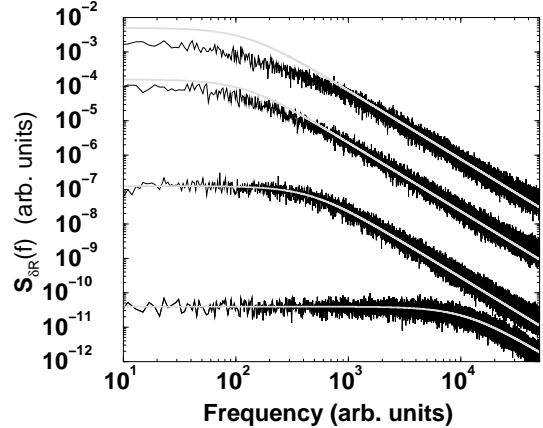,width=7.0cm}
   \caption{Power spectral density of resistance fluctuations of 
    $125\times 125$ networks for increasing  $E_R$ values. Precisely, starting 
    from the bottom: $E_R=0.026, 0.103, 0.155, 0.164$ eV. In all cases the 
    current corresponds to the respective threshold for breakdown 
    (respectively: $I=4.4, 1.15, 0.14, 0.011$ A). For visual reasons the 
    lowest spectrum has been shifted down by multipling it by a factor 
    $5 \times 10^{-3}$. The gray lines represent the best-fit with a 
    Lorentzian spectral density.}
    {\label{fig:7}}
    \end{figure} 

Other interesting information about the system can be extracted from the 
spectral analysis  of resistance fluctuations \cite{weissman,antal,racz}.
Hence we report in Fig. 7 the power spectral density of resistance 
fluctuations of $125 \times 125$ networks at increasing values of $E_R$.
All the spectra are calculated for networks stressed by currents corresponding
to the threshold for electrical breakdown. Starting from the bottom, the 
curves are obtained  respectively for $E_R=0.026, 0.103, 0.155$ and 
$0.164$ eV (the same values of $E_R$ considered in Fig. 5). Basically all
the spectra display a Lorentzian behavior which witnesses an exponential decay 
of the auto-correlation function of resistance fluctuations \cite{weissman}
(the gray curves in this figure represent the best-fit with a Lorentzian 
distribution). Moreover, we can see that for increasing values of $E_R$, 
i.e. for increasing level of intrinsic disorder, the corner frequency of 
the spectra is systematically shifted towards lower values. This occurs 
because, at increasing values of $<p>$, the network approaches the 
percolation threshold, thus the correlation length increases as 
$\xi \sim |<p>-p_c|^{-\nu}$ (where $\nu$ is the correlation length exponent) 
\cite{stauffer} and the same occurs for the correlation time of resistance 
fluctuations. However, we note that for the two upper spectra in Fig. 7 
the fit with a Lorentzian curve appears progressively less satisfactory in the
low frequency region. On the other hand, these spectra are obtained by taking 
the value of $E_R$ near or very near to $E_{R,MAX}$ and thus correspond to
networks progressively closer to the critical conditions. This anomalous
behavior in the low frequency region of the power specral density of resistance
fluctuations in the case of networks close to criticality, could even suggest 
the emergence of a $1/f$ behavior. Longer time series are required to point
out the behavior of the spectra in this frequency region. 

\section{CONCLUSIONS}

We have studied the distribution of the resistance fluctuations of conducting 
thin films with different levels of internal disorder. The study has been 
performed by describing the film as a resistor network in a steady state
determined by the competition of two biased stochastic processes, according to 
the SBRN model \cite{upon99,SBRN_1,SBRN_2}. We have considered systems of
different sizes and under different stress conditions, from the linear 
response regime up to the threshold for electrical breakdown. A remarkable
non-Gaussianity of the fluctuation distribution is found near breakdown. 
This non-Gaussianity becomes more evident at increasing the degree of disorder
of the network. As a general trend, these deviations from Gaussianity are 
related to the finite size of the system and they are found to vanish in the 
large size limit. However, near the critical point of the conductor-insulator 
transition, the non-Gaussianity is found to persist in the large 
size limit and is well fitted by the universal Bramwell-Holdsworth-Pinton 
distribution \cite{bramwell_nat,bramwell_prl}. Furthermore, in the last case, 
the non-Gaussianity seems to be associated with an anomalous behavior in the 
low frequency region of the specral density of resistance fluctuations. 

\section*{ACKNOWLEDGEMENTS}
Partial support from the cofin-03 project ``Modelli e misure di
rumore in nanostrutture'' financed by Italian MIUR and from the SPOT-NOSED  
project IST-2001-38899 of EC is gratefully acknowledged.

\end{multicols}

\end{document}